\documentclass[letterpaper]{article}

\usepackage{emulateapj}
\usepackage{onecolfloat}
\usepackage{apjfonts}
\usepackage{amsmath}
\usepackage{natbib}

\usepackage{graphicx}
\usepackage[figuresright]{rotating} 
\usepackage{float}       
\usepackage{afterpage}   

\newcommand{\rxte}{\emph{RXTE}}
\newcommand{\wsim}{\ensuremath{\sim}}
\newcommand{\xf}{XTE~J1550$-$564}
\newcommand{\xs}{XTE~J1650$-$500}

\newcommand{\gx}{GX~339$-$4}
\newcommand{\fu}{4U~1630$-$47}
\newcommand{\reso}{XTE~J1748$-$288}
\newcommand{\maso}{XTE~J1859$+$226}
\newcommand{\ind}{$\Gamma$}
\newcommand{\tin}{$T_{in}$}

\begin{document}


\twocolumn[
\title{A close look at the state transitions of Galactic black hole transients during outburst decay}

\author{E. Kalemci\altaffilmark{1},
 J. A. Tomsick\altaffilmark{2},
 R. E. Rothschild\altaffilmark{2},
 K. Pottschmidt\altaffilmark{3,4},
 P. Kaaret\altaffilmark{5}
}

\altaffiltext{1}{Space Sciences Laboratory, 7 Gauss Way, University of 
California, Berkeley, CA, 94720-7450, USA}

\altaffiltext{2}{Center for Astrophysics and Space Sciences, Code
0424, University of California at San Diego, La Jolla, CA,
92093-0424, USA}

\altaffiltext{3}{Max-Planck-Institut f\"ur Extraterrestrische Physik, Giessenbachstr. 1, 85748 Garching, Germany}

\altaffiltext{4}{\emph{INTEGRAL} Science Data Centre, Chemin d'\'Ecogia 16, 1290 Versoix, Switzerland}

\altaffiltext{5}{Harvard-Smithsonian Center for Astrophysics, 60 Garden Street,
 Cambridge, CA, 02138, USA}


\begin{abstract}

We characterize the evolution of spectral and temporal properties of several 
Galactic black hole transients during outburst decay using the data from
well sampled PCA/\rxte\ observations close to the transition to the low/hard 
state. We find several global patterns of evolution for spectral and temporal
 parameters before, during, and after the transition. We show that the changes
in temporal properties (sudden increase or decrease in the rms amplitude of 
variability) are much sharper than the changes in the spectral properties, and
it is much easier to identify a state transition with the temporal properties.
 The spectral index shows a drop 3-5 days before the transition for some of our 
sources. The ratio of the power-law flux to the total flux in the 3-25 keV band 
increases close to the transition, which may mean that the system must be 
dominated by the coronal emission for the transition to occur. We also 
show that the power-law flux shows a sharp change along with the temporal 
properties during the transitions which may indicate a threshold transition 
volume for the corona. The evolution of the spectral and temporal properties 
after the transition is consistent with the idea that the inner accretion disk 
moves away from the black hole. Based on the evolution of spectral and 
temporal parameters and changes during the transitions, we discuss possible 
scenarios of how the transition is happening.

\end{abstract}


] 


\section{Introduction}\label{sec:intro}

X-ray observations of Galactic black holes (GBH) indicate that they are found 
in several distinct spectral states \citep{Tanaka95,McClintock03}. These 
states are generally determined by the relative strength of two different 
emission components: soft blackbody-like radiation from an optically thick,
 geometrically thin accretion disk and a harder component showing a power-law 
spectrum believed to originate from Compton upscattering of soft seed photons 
from the disk by a hot electron corona. Often a relation between the X-ray 
luminosity of the source and spectral states also exists. When the soft 
component dominates the spectrum, the 2--10 keV luminosity is relatively high
 ($\gtrsim 10^{37}$ ergs/s), and therefore this state is called the 
``high/soft state'' (HS). When the hard component dominates, then usually the 
2--10 keV luminosity is low ($\lesssim 10^{37}$ ergs/s), and this state is 
called the ``low/hard state'' (LS). This dependence shows that the mass 
accretion rate plays an important role in determining the spectral states. 
These states also differ in terms of their short-term timing properties. The 
HS is often characterized by lack of or a very low level of variability, 
whereas the LS shows very strong variability (\wsim 30\% rms in 0.04--4 Hz).
 Recently, changes in the radio properties are also associated with the 
spectral states. In the LS, compact, optically thick jets are observed
 \citep{Fender01}. During state transitions optically thin outflows are 
detected \citep{Corbel01}, and during the HS the radio emission is quenched 
\citep{Fender01}. There are also cases for which both spectral components are
 present at comparable strength, characterized by parameters that are 
intermediate between those of the HS and the LS. For such a case, if the 
source flux is lower than the HS flux, the state is often called the 
``intermediate state'' (IS), and if the flux is higher than the HS flux, it is
 called the ``very high state'' (VHS). We note that the classification of 
these states is not rigorously defined and is still an active topic of debate. 

\begin{figure}[t]
\plotone{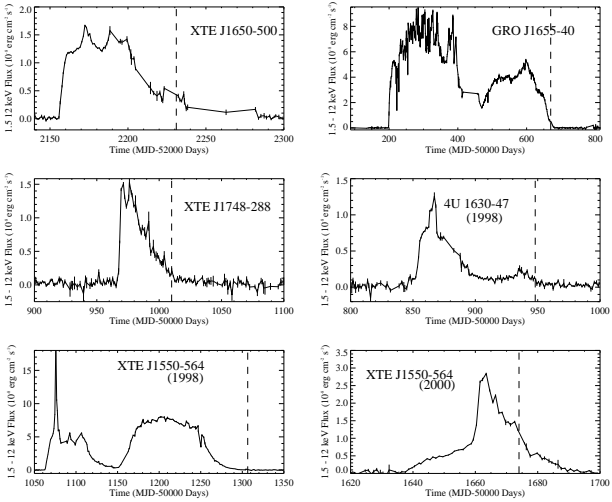}
\caption{\label{fig:asm_many}
The \rxte, All Sky Monitor light curves of six outbursts in 1.5 - 12 keV band. The 
dashed lines represent the time of transition to the LS during outburst decay.
}
\end{figure}

Most of the GBHs are observed during outbursts caused by a sudden, dramatic 
increase of mass accretion rate onto the black hole. Fig.~\ref{fig:asm_many} shows 
several examples of outburst light curves discussed in this report (see 
\citealt{Kalemci_tez} for all of the ASM light curves.). Because of the
 dependence of spectral states on the mass accretion rate, a GBH transient often 
follows a specific sequence of spectral states. It is usually observed in the LS at 
the beginning of the outburst. As flux increases, it makes a transition to the HS or 
the VHS. As the source decays towards quiescence, a transition to the LS is usually 
observed. Some transients might follow a more complicated sequence of states, and 
some stay in the LS throughout the outburst. It is generally believed that the state 
transitions involve large restructuring of the accretion geometry of 
GBHs \citep{Esin97,Zdziarski02_2}. Therefore, analysis of these sources during
state transitions may probe the dynamics of their accretion structure. Although the
 mass accretion rate is a very important parameter determining the spectral states,
 it is unlikely that the states and transitions are solely determined by this 
parameter. Some sources show hysteresis of the transition luminosities \citep[the 
luminosity of transition from LS to HS during the beginning of the outburst is much
 higher than the luminosity of the transition from HS to LS during the outburst 
decay,][]{Miyamoto95,Nowak02b}, and for some sources, a second, independent 
parameter seems to be required to explain the complexity of transitions. It is not
 clear what this second parameter is. It can be the position of the inner edge of 
the accretion disk, but this parameter may not be completely independent from the 
mass accretion rate (e.g. \citealt{Esin97,Meyer00}). Based on the behavior of the 
1998 outburst of \xf, \cite{Homan01} claimed that the second independent parameter 
may be the size of the corona. The transitions may also be a result of an overall 
change in the type and the geometry of the corona. \cite{Zdziarski02_2} explains 
the different states in Cyg~X-1 by the change of an accretion structure which 
consists of a hot inner accretion flow surrounded by an optically thick disk 
truncated far away from the minimum stable orbit in the LS, to one that 
consists of flares and active regions above an accretion disk extending close 
to the minimum stable orbit in the so-called ``soft state''. We note that 
for Cyg X-1, the soft state is more similar to the IS of GBH transients rather than
 the HS. According to the hybrid model of \cite{Coppi00}, the state transitions 
may be explained by a change of size and electron energy distribution of 
the corona. In the LS, the inner part of the disk puffs up and becomes the 
hot corona, dominated by thermal electrons. Non-thermal electrons may also be 
present. In the HS, the edge of the cool accretion disk is close to the minimum 
stable orbit, and a small, non-thermal corona is also present. In this case, Compton 
scattering of disk photons by the non-thermal electrons is responsible for the steep 
power-law observed in the spectra of GBHs in the HS. It has also been claimed that 
this steep power-law component is a manifestation of Compton up-scattering of disk 
photons by a converging inflow of material inside the last stable orbit 
\citep[][and references therein]{Laurent01}. Then the transition might be related to 
the type of Comptonization: bulk motion in the HS versus thermal (or hybrid) in the 
LS. 

The transition to the LS not only results in sharp changes in the X-ray 
spectrum, but also creates a physical environment that shows strong broad-band 
variability and quasi-periodic oscillations (QPO). The changes in the spectral 
parameters close to the state transition might provide important clues to 
understand how this transition is occurring and the driving force of the 
observed variability. Our group has been observing GBH transients during 
outburst decays in X-rays with the \emph{Rossi X-ray Timing Explorer} (\rxte) 
and in radio. We have quantified the evolution before, during and after the 
transition to the LS for various sources 
\citep{Kalemci01,Tomsick01b,Kalemci02,Tomsick03_2}. In 
Section~\ref{subsec:evolto}, we combine results of spectral and temporal 
analyses of the \rxte\ data from individual sources during outburst decay 
using our observations as well as the archival data, and investigate the 
evolution of the spectral index (\ind), the inner disk temperature (\tin), the 
power-law flux, the disk-blackbody (diskbb) flux and the power-law fraction (PLR, 
ratio of the power-law flux to the total flux in 3--25 keV band) to obtain a global 
understanding of the physical environment before and during the state transition.

Another important topic is the evolution of spectral parameters after the
state transition, which provides an understanding of the dynamics of the accretion 
systems in the LS. In addition to the spectral evolution, the evolution of the 
temporal parameters is another valuable tool in the LS. Thus, in 
Section~\ref{subsec:evolafter}, we analyze the evolution of spectral and temporal 
parameters of GBH transients after the state transition. The temporal parameters we 
use are the characteristic frequencies of the Lorentzian components in the 
power-spectral-density (PSD) fits and the rms amplitude of variability.
 

\section{Observations and Analysis}\label{sec:obs}

We analyzed the PCA/\emph{RXTE} (Proportional Counter Array, see 
\citealt{Bradt93} for a description of \emph{RXTE}) data from all GBH 
transients that have been observed with \emph{RXTE} between 1996 and 2001 that 
made a state transition during outburst decay. Eight sources in eleven 
outburst decays obey these source selection criteria. These sources and 
outburst years are : \xs\ in 2001, GRO~J1655$-$40 in 1996, \reso\ in 1998, 
XTE~J1755$-$324 in 1997, \gx\ in 1998 (decayed in 1999), \fu\ in 1998, 1999 
and 2001, \xf\ in 1998 and 2000 and \maso\ in 1999. Although XTE~J1755$-$324 obeys 
the criteria, it has very poor coverage and results from this source are 
not included in this work. This decreases the number of sources to seven, and the 
number of outbursts to ten. Note that \maso\ did not make a ``traditional'' 
transition to the low/hard state, however it showed timing noise for some 
observations during the decay (see \citealt{Kalemci_tez} and references 
therein for the properties and observation times of each 
source).

For all of the spectral analysis, we fit the data in the 3--25 keV band using 
the response matrix and the background model created using the standard FTOOLS 
(version 5.2) programs. We added a 1\% systematic error to the spectra to 
account for uncertainties in the PCA response. We used a 
multi-component spectral model consisting of a power-law, a multi-color 
disk blackbody \citep[\textbf{diskbb} in XSPEC, ][]{Makishima86}, a broad 
absorption edge \citep[\textbf{smedge} in XSPEC, ][]{Ebisawa94} with 
interstellar absorption (\textbf{phabs} in XSPEC). This model has been commonly
used for the spectral analysis of GBHs in the LS 
\citep{Tomsick00,Sobczak00}. For all observations, the reduced $\chi^{2}$ is 
between 0.5 and 1.5. For some observations, in order to reach acceptable 
$\chi^{2}$ values, a Gaussian iron line feature was needed. However, the 
parameters of the iron line have not been used in the analysis. For some of the
outbursts, we used published spectral fit parameters if the same model was applied 
for the fit (\citealt{Tomsick00} for \fu\ in 1998, and
\citealt{Tomsick01b} for \xf\ in 2000). All PCA fluxes in this report are unabsorbed 
model fluxes to remove source to source variations due to different absorption 
column densities. The details of spectral fit models and parameters for each source 
can be found in \cite{Kalemci_tez}. 

For the temporal analysis, we were as uniform as possible in terms of 
choosing energy bands, time resolution and segmentation of light curves for 
all of our sources. However, since some of this work is based on the 
analysis of the archival data, we were limited to the choice of data modes by 
the PI of the original proposal. For most of the observations we used 2--26 keV
 energy band, a Nyquist frequency of 256 Hz and 256 s light curve segments.
 Although there may be slight differences from source to source in terms of 
energy band, and highest and lowest PSD frequencies, these differences are not 
critical since we look for trends rather than absolute values.

\begin{figure}
\plotone{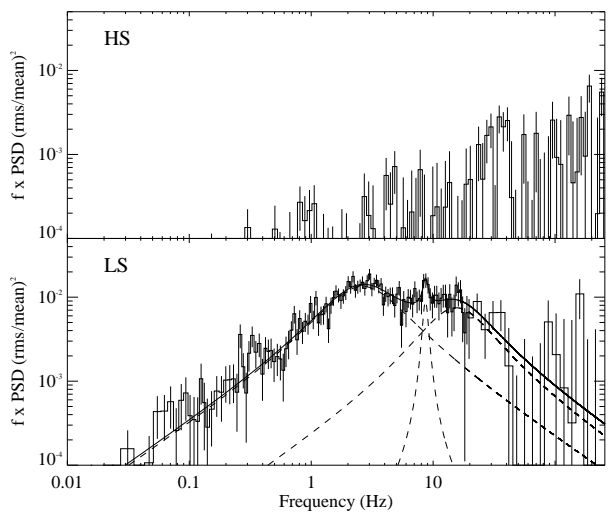}
\caption{\label{fig:lorex}
Power spectra of \xs\ in the HS (top panel) and in the LS (bottom panel). The PSD
in the LS is fitted with two broad and a narrow Lorentzian. The HS observation has
a 2$\sigma$ upper limit of 4\% on the rms amplitude of variability, whereas the LS 
observation 2 days later has an rms amplitude of variability of $12.43 \pm 0.55$ \%}
\end{figure}

   Historically, the PSD of GBHs during the low state has been modeled by a 
broken power-law (or power-laws with more than one break) plus narrow 
Lorentzians to fit the QPOs \citep{Nowak99,Tomsick00,Kalemci01}. However, 
recent papers successfully fit several GBH and neutron star PSDs with broad 
Lorentzians for the continuum and narrow Lorentzians for the QPOs 
\citep{Belloni02,vanStraaten01,Pottschmidt02,Kalemci02}. Following this
approach, we fit all our PSDs with Lorentzians of the form:
\begin{equation}
L_{i}(f)\;=\;{{R_{i}^{2} \; \Delta_{i}}\over{2 \, \pi \; [(f-f_{i})^{2}+({1\over{2}}\,\Delta_{i})^2]}}
\end{equation}
where subscript $i$ denotes each Lorentzian component in the fit, $R_{i}$ is 
the rms amplitude of the Lorentzian in the frequency band of -$\infty$ to 
+$\infty$, $\Delta_{i}$ is the 
full-width-half-maximum, and $f_{i}$ is the resonance frequency. A useful
quantity of the Lorentzian is the ``peak frequency'' at which the Lorentzian
 contributes maximum power per logarithmic frequency interval:
\begin{equation}
\nu_{i}\;=\;{f_{i}\;\left({{\Delta_{i}^{2}}\over{4 \, f_{i}^{2}}}+1\right)^{1/2}}
\end{equation}
Fig.~\ref{fig:lorex} shows an example LS power spectrum of \xs\ in the form of {PSD 
$\times$ frequency}, along with broad and narrow Lorentzians fit components. In this 
figure, the Lorentzians peak at $\nu_{i}$, demonstrating the easy identification of 
characteristic frequencies as peak frequencies of Lorentzian components. Some of our 
observations contain a Lorentzian that is narrow (with quality value 
$Q_{i} = f_{i}/{\Delta_{i}}>2$, as compared to $Q<1$ for broad Lorentzians) which we 
call a QPO. In this work, for each observation, the term ``characteristic 
frequency'' represents the resonance frequency of the fundamental QPO\footnote{The 
peak frequency and the resonance frequency differ by $<$3\% for QPOs with $Q>$2.} if 
present, and otherwise the lowest peak frequency of the broad Lorentzian components 
in the PSD\footnote{The lowest peak frequency is nearly equivalent to the 
``break frequency'' if broken power-law modeling is adopted, see \cite{Belloni02} 
for a detailed discussion.}. Note that since the QPO frequencies and the Lorentzian 
peak frequencies are all shown to be correlated \citep{Belloni02}, it does not 
matter which one we use to characterize the evolution. The rms amplitudes are 
calculated over a frequency band from zero to infinity.

 Mostly due to our group's monitoring program of GBHs with \rxte\ during 
outburst decay, we obtained very good coverage for 3 sources, in 6 different 
outbursts (1998, 1999 and 2001 outbursts of \fu, 1998 and 2000 outbursts of
 \xf, and the 2001 outburst of \xs\ with almost daily monitoring). 
 GRO~J1655$-$40 and \gx\ are excluded from the discussion of the changes before
and during the transition in Section~\ref{subsec:evolto} since they have poor 
coverage ($\geq$10 days between observations). XTE~J1748$-$288 with 
observations every \wsim5 days is included in the discussion of the evolution 
to the state transition, however is not included in the discussion of the 
changes during the transition. Results from these sources are included in 
the discussion of the evolution after the transition for which very good coverage 
is not necessary. All the data used in this report are available on-line as a 
machine-readable table.


\section{Results}\label{sec:results}

\begin{figure}
\plotone{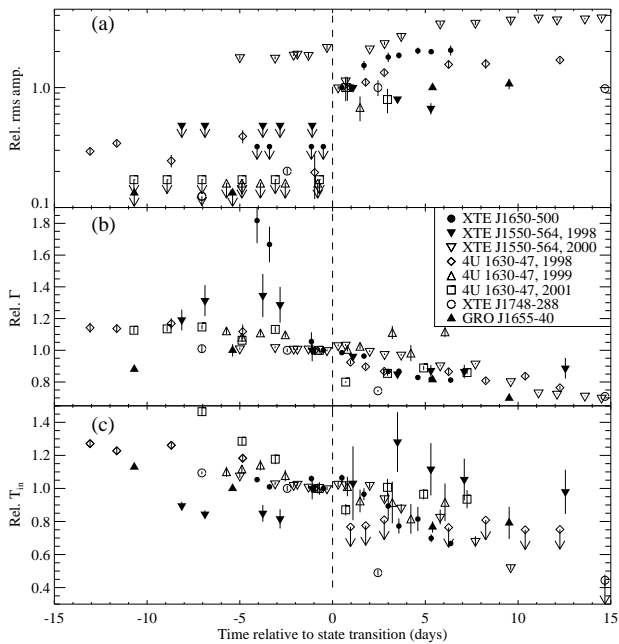}
\caption{\label{fig:evolmix1}
Evolution of (a) the rms amplitude of variability, (b) the spectral index, and
 (c) the inner disk temperature. The state transition is assumed to have happened 
exactly in between the observations closest to the sharp change observed in 
panel (a), and represented by a dashed line. For (a), the values for each 
source are normalized with respect to the value just after the state 
transition. For both (b) and (c), the values for each source are normalized 
with respect to the value just before the state transition. For some points, the
1 $\sigma$ errors are smaller than the plot symbols. The upper limits are 2$\sigma$.}
\end{figure}

In this work, we define the state transition in terms of sharp changes in 
variability properties rather than sharp changes in spectral properties. All 
of our sources with good coverage showed sharp, distinct changes in terms of 
variability in less than two days, but this was not the case for the spectral 
properties. Most of the transitions occurred from the HS to LS, and for those 
cases the transition was marked by a very large increase in the total rms 
amplitude of variability as shown in Fig.~\ref{fig:evolmix1}a. For most of the 
cases, the transition was from a featureless, Poisson noise dominated PSD with 
only a few \% rms amplitude upper limit to a PSD showing well defined broad-band 
variability. In other words, the variability ``appeared'' on a few days timescale 
for these systems. Fig.~\ref{fig:lorex}  represents this appearance of variability, 
as the observation in the HS (top panel, with $<$4\% rms upper limit variability) 
and the observation in the LS (bottom panel, with $>$12\% rms variability with well
defined Lorentzian components and a QPO) in this figure are only two days apart.
 Since the HS PSD is often featureless and has very low or no rms amplitude of 
variability, it is not possible to discuss the evolution of characteristic 
frequencies before the transition for none of our sources except the 2000 outburst 
of \xf, which showed a complex pattern. It was in the IS, with rms amplitude 
variability of \wsim13\%. During the IS, it showed a QPO with constant frequency at 
\wsim 9 Hz. Then it showed a large drop in variability to an amplitude of 
\wsim7\%. We marked this change as the transition. Two days later, the rms amplitude 
of variability jumped again to levels of 15\%. The morphology of the PSDs was 
different in the IS, LS, and during the time that the rms amplitude showed a drop 
\citep{Kalemci_tez}. Despite the complexity of some individual cases, it is possible 
to infer global patterns of evolution for different parameters before, during and 
after the state transition.

\subsection{Changes in spectral properties that lead to state transitions}\label{subsec:evolto}

\begin{figure}
\plotone{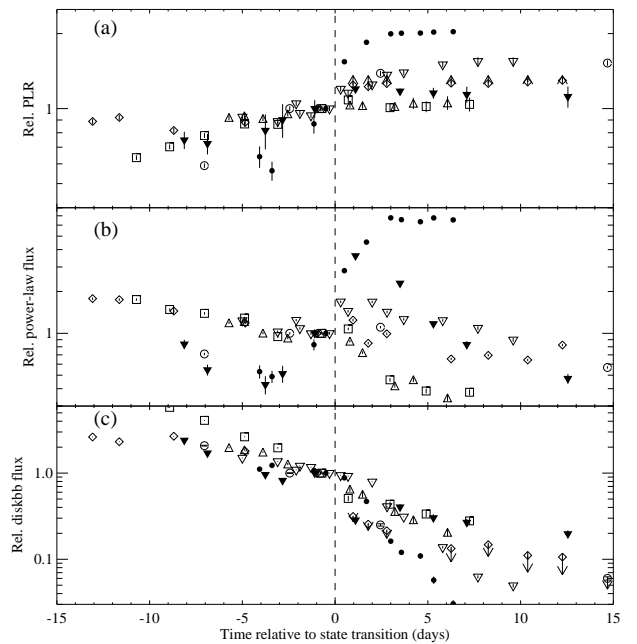}
\caption{\label{fig:evolmix2}
Evolution of (a) the PLR, (b) the power-law flux, and (c) the diskbb flux. The
 dashed line represents the time of transition. The values for each source are
 normalized with respect to the value just before the state transition. A legend 
is given in Fig.~\ref{fig:evolmix1}.  For some points, the 1 $\sigma$ errors are 
smaller than the plot symbols. The upper limits are 2$\sigma$}
\end{figure}

\begin{figure}[t]
\plotone{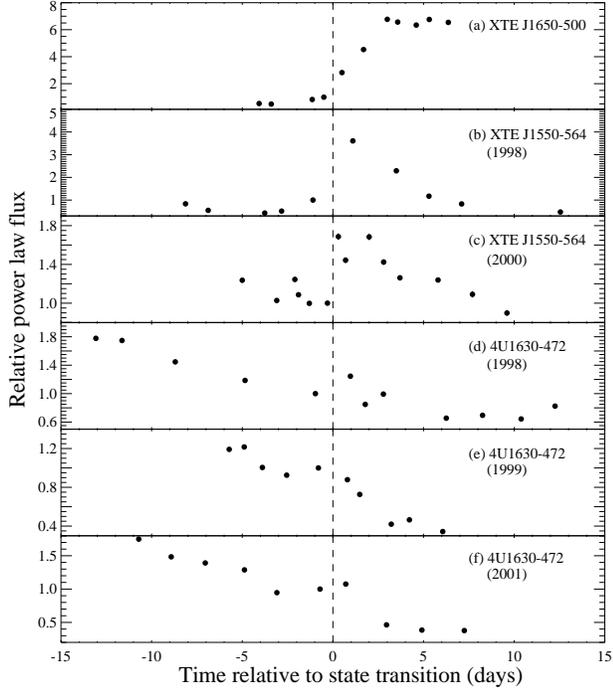}
\caption{\label{fig:evolplf}
Evolution of the power-law flux before and after the transition. The dashed 
line indicates the time of state transition. The values for each source are 
normalized with respect to the value just before the state transition. The 1 
$\sigma$ errors are smaller than the plot symbols
}
\end{figure}

First, we characterize the changes in spectral properties before the state
transition. The evolution of the photon spectral index (\ind), and the inner disk 
temperature (\tin) close to the transition are shown in Fig.~\ref{fig:evolmix1}b 
and c respectively. The evolution of the PLR, power-law flux and the diskbb flux 
are shown in Fig.~\ref{fig:evolmix2}. In most sources, \ind\ 
(Fig.~\ref{fig:evolmix1}b) is approximately constant until 3-5 days before the 
timing transition at which point it begins to decline. For some sources, there is 
a distinct drop in \ind\ at the beginning of the decline. Except for the 1998 
outburst of \xf, \tin\ (Fig.~\ref{fig:evolmix1}c) either decreases or stays
 constant before the transition. The power-law flux (Fig.~\ref{fig:evolmix2}b) 
shows a complex behavior (see also Fig.~\ref{fig:evolplf}), and the 
diskbb-flux (Fig.~\ref{fig:evolmix2}c) decreases for majority of the outbursts, or 
stays constant. The cumulative effect of the power-law and diskbb-flux evolution 
for the majority of the outbursts is an increasing PLR (Fig.~\ref{fig:evolmix2}a) 
before the transition. 

We established that the transition occurs in a few days time scale for GBH 
transients. The change in variability properties (usually, the appearance of
variability) that defines the transition should be a response to the change of 
one or more spectral parameters. To understand which parameter drives the 
transitions and appearance of variability, we compared the two observations 
just before and after the transition (observations just before and  
after the dashed lines in Figs.~\ref{fig:evolmix1}~and~\ref{fig:evolmix2}). 
We specify that a spectral parameter is showing a sharp change during the 
transition if either the slope of the parameter as a function of time changes
 sign or the percentage change in the value of the parameter compared to the 
previous observation is at least three times that of the previous observation.
 With this definition, two parameters show a sharp change for the majority of 
the outbursts: the power-law flux (see Fig.~\ref{fig:evolplf}) and the 
PLR\footnote{\gx\ during its recent decay from the outburst showed a large 
increase in the PLR, and by using this information, we were able to estimate 
when the transition would occur and scheduled a successful \rxte\ observation 
to catch the beginning of the transition.}. The PLR is not an independent 
parameter, and in this case its increase is mostly driven by the increase in the 
power-law flux. For both the 1999 and the 2001 outbursts of \fu, there is no 
sharp change in any of the spectral parameters between those two observations,
 yet one observation shows no variability and the next one shows variability. 
However, for both, the evolution of the power-law flux shows a change in the slope 
an observation earlier. The situation is more complicated for the 2000 outburst 
of \xf. There is a large increase in the power-law flux and PLR during the 
transition, but this change corresponds to a decrease in the rms amplitude 
\citep{Kalemci_tez}. 

Analysis of PCA data from GBH transient \maso\ during its outburst 
decay revealed interesting behavior in terms of variability and spectral 
evolution and provided another case for the relation between the power-law 
flux and variability. \maso\ did not show a traditional transition, but showed 
variability for some of the observations during the decay. In 1 day 
timescale, the power-law flux almost doubled and variability appeared, and 
when the power-law flux dropped below a threshold value, variability disappeared. 
Except the PLR, which is tied to the power-law flux, no other spectral 
parameter showed a sharp change between observations that show variability, 
and those that do not show variability for this source 
\citep{Kalemci_tez}.

\subsubsection{A discussion of transition fluxes}

\begin{figure}[t]
\plotone{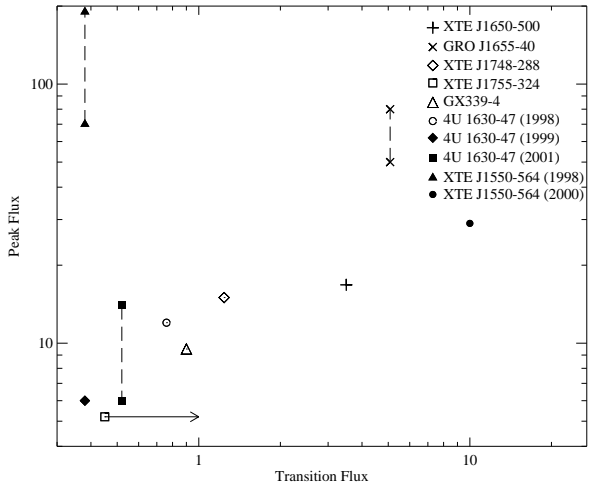}
\caption{\label{fig:trflux}
Peak flux versus transition flux in 1.5 -- 12 keV band. Fluxes are in units of
$\rm 10^{-9}\, ergs \, cm^{-2} \, s^{-1}$. For the points connected with dashed
 lines, both peak fluxes and plateau fluxes are shown.
}
\end{figure}

It is important to get an idea of when the transition happens during
the decay for effective monitoring of the sources with pointed instruments. 
Often the sources are not monitored frequently enough (i.e., once a day) with 
pointed observations, but the ASM measures the flux in 3 energy bands very 
frequently, which provides an opportunity to detect transitions without 
re-pointing the satellite. (See Fig.~\ref{fig:asm_many} for the ASM light curves
of some of our sources.) The transition is often detected by a sudden 
increase in the hardness ratio 2 (HR2, the ratio of 5--12 keV flux to 1.5--3 
keV flux). A few days just before and after the transition may manifest the 
most interesting behavior. A relationship between the peak flux and the 
transition flux (1.5 -- 12 keV flux for the observation that shows 
variability) had been realized earlier (Tomsick, private communication). We 
plot this relation for all of the outbursts that show a state transition in
 Fig.~\ref{fig:trflux}. Except the 1998 outburst of \xf, which showed one of 
the strongest flares in the \rxte\ history, the transition flux increases with 
the peak flux. The Spearman's rank order correlation coefficient for this data
 set (excluding the 1998 outburst of \xf) is 0.90, pointing out a correlation. 
On the other hand, the linear correlation coefficient is 0.52, indicating 
a relation which is not strictly linear.

Some outbursts have the usual ``fast-rise-exponential-decay'' (FRED) shape 
(such as \reso), and some have a very complicated shape (such as \fu\ in 2001).
 The complicated ones may show a plateau before the final decay (like GRO~J1655$-$40,
and \xf\ in 2000 in Fig.~\ref{fig:asm_many}). For these sources, we also plotted the 
plateau ASM flux along with the peak flux. Replacing the peak flux by the plateau 
flux for these sources improves the linear correlation for the whole sample to $r$ of
 0.706. The linear correlation is not very strong, but it allows for an estimate of 
when the transition might happen, which is useful for planning observations to study 
the transition.

\subsection{Evolution of spectral and temporal parameters after the transition}\label{subsec:evolafter}

\begin{figure}[t]
\plotone{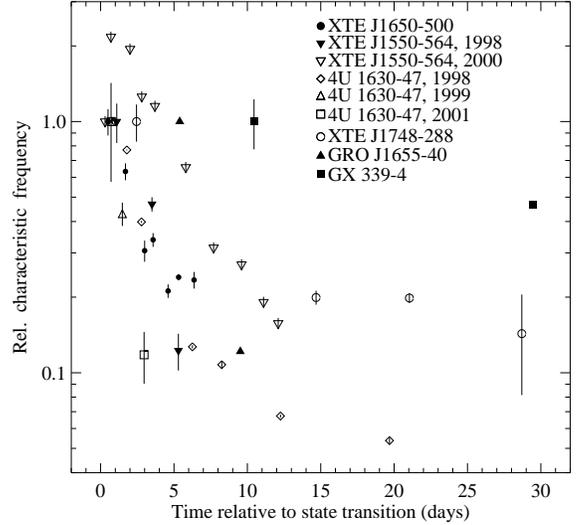}
\caption{\label{fig:evolcf}
Evolution of the characteristic frequencies (the QPO resonance frequency for
\xf\ in 2000, \fu\ in 1998 and 1999, and \reso, lowest peak frequency of the 
broad Lorentzian components in the PSD for the remaining) after the state 
transition. For some points, the 1 $\sigma$ errors are smaller than the plot 
symbols.
}
\end{figure}

 The evolution of spectral and temporal parameters after the transition is 
shown in Figs.~\ref{fig:evolmix1},~\ref{fig:evolmix2},~\ref{fig:evolplf},~and~
\ref{fig:evolcf}. For \xs, \gx\ (not shown in 
Figs.~\ref{fig:evolmix1},~\ref{fig:evolmix2},~\ref{fig:evolplf} because of 
poor coverage), \fu\ in 1998 and \xf\ in 2000, \ind\ decreases after the 
transition. For other outbursts, it is either constant or shows irregular behavior.
 The power-law flux decreases for all of the outbursts except \xs\ 
(see Figs.~\ref{fig:evolmix2}~and~\ref{fig:evolplf}). For most cases, 
the diskbb-flux and \tin\ either decrease, or are unobservable after the 
transition. It is hard to constrain the evolution of \tin\ for the 1998 
outburst of \xf\ and the 2001 outburst of \fu.  For all outbursts, the diskbb 
component is unobservable within 15 days of the state transition. For \xs\ and the 
2000 outburst of \xf, the PLR first increases and then stays constant. For all other 
outbursts, the PLR is very close to unity and does not vary. 

 The evolution of the characteristic frequencies after the transition is shown
in  Fig.~\ref{fig:evolcf}. Except a single observation at the beginning of the
2000 outburst of \xf, all characteristic frequencies decrease after the 
transition. There are only two observations that show variability in the 1999 
and 2001 outbursts of \fu, and even for those cases, the later observations 
have smaller characteristic frequencies. The behavior of the rms 
amplitudes are more complicated (see Fig.~\ref{fig:evolmix1}a). The 1998 
outburst of \xf\ and \reso\ show a decreasing trend. The rms amplitudes for 
\xs, the 2000 outburst of \xf\ and the 1998 outburst of \fu\ increase and
 level off. For GRO~J1655$-$40, the rms amplitudes are consistent with being 
constant. For \gx\ on the other hand, the rms amplitude increases 
\cite{Kalemci_tez}. 


\section{Discussion}

\subsection{State transition and appearance of variability}

\begin{figure}
\plotone{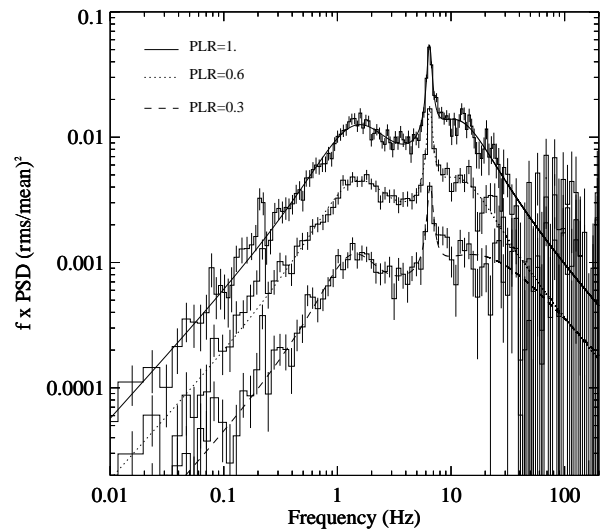}
\caption{\label{fig:simmulplot}
PSD of GRO~J1655$-$40 during an observation in the LS is shown with the solid 
curve obtained by fitting the PSD with two broad Lorentzians and a QPO. The 
dotted and the dashed curves are fits to the PSDs when Poisson noise is added 
to the original light curve. The PSD under the dotted curve corresponds to a PLR 
of 0.6 and the PSD under the dashed curve corresponds to a PLR of 0.3. 
}
\end{figure}

 First, we discuss possible reasons for the state transition and appearance of
 variability based on our results above. Suppose almost all of the 
variability we observe is due to the photons coming from the power-law
component presumably originating in the corona, and the disk photons are 
providing mostly Poisson noise. This idea is supported by the fact that 
very little or no variability is observed during the HS when the diskbb 
dominates, and strong variability is observed when the power-law component 
dominates in the LS. The anti-correlation between the diskbb flux and the rms 
amplitude observed both in \xs\ \citep{Kalemci02} and the 2000 outburst of 
\xf\ \citep{Kalemci01} is also potentially consistent with this scenario. It
 is conceivable that the Poisson noise dominated diskbb flux reduces the rms 
amplitudes of variability to unobservable levels. Then the PLR would be a good 
indicator of when the variability will be observed, since the higher the PLR, the
lower the diskbb flux compared to the total flux. For all of our sources, no 
variability is observed if the PLR$<$0.45. However, the value of the PLR cannot be
 the sole determinant of when the variability will be observed. For most of 
the sources the PLR has to be greater than 0.8 for the variability to appear, 
although it can be as low as 0.45 for \maso\ \citep{Kalemci_tez}. To test the 
argument that the diskbb flux might reduce the rms amplitudes to unobservable 
levels, we conducted very simple simulations. We added various levels of 
Poisson noise to the light curve of an observation of GRO~J1655$-$40 taken on 
August 14, 1997, and compared the resultant PSD with the original PSD. We chose 
this observation since it has a complex structure, and no diskbb flux, so all the 
emission is coming from the corona. The results are shown in 
Fig.~\ref{fig:simmulplot}. Although this figure clearly shows that the rms 
amplitude of the PSD depends on the PLR, it also shows that even for a PLR of 0.3,
 the variability is clearly observed. There is no large change in the PSD shape. 
Moreover, lack of diskbb emission \textbf{does not} guarantee broad-band 
variability even if a hard component exists. There is strong emission in the 
6--15 keV band for most of the sources before the transition, and spectral analysis
 shows that only a few percent of the emission in this band is from the soft 
component. Nonetheless, although completely power-law dominated, these 
observations do not show variability. Note that, these simple simulations do not 
take into account the effects of strong disk emission on the corona. The two 
emission components are not independent of each other, because Compton cooling of 
the corona by the diskbb emission can affect the temperature and density structure 
of the corona. The amount of diskbb radiation cannot be ruled out as an important 
parameter in determining when the variability will be observed.

The power-law emission dominates the LS when the variability is observed but 
its presence alone is not enough to create broad-band variability. These 
findings suggest several possibilities for the presence of variability in 
the LS. The origin of the power-law component in the LS and the other
states might be completely different. The passage from the IS to the LS (as 
observed for \gx\ and the 2000 outburst of \xf) might be due to a change of the 
form of the corona from active regions above the accretion disk to an inner 
accretion flow \citep{Zdziarski02_2}. The power-law component in the HS 
may be a result of Comptonization of disk photons by non-thermal 
electrons \citep{Coppi00}, or bulk motion Comptonization \citep{Laurent01},
whereas in the LS, it may be due to thermal Comptonization. However, if the form 
or the composition of the corona is changing during the transition, one 
would expect a sharp change not only in power-law flux, but also in \ind, especially 
if the change is from bulk motion Comptonization to thermal Comptonization. There 
are two ways to increase the power-law flux: increasing the number of seed photons; 
and/or increasing the area of the corona that intercepts the seed photons. It is 
unlikely that the number of seed photons are increasing, the corona is believed to be
 optically thin, and an increase in the soft flux should be observed (unless most 
photons have energies below the PCA range). Therefore, our findings are consistent 
with the idea that, during the transitions, the size of the corona increases to a 
threshold value for variability, which would be consistent with the idea that the 
second independent parameter determining the spectral states is the size of the 
Comptonizing region \citep{Homan01}. The strongest support for this explanation 
comes from the behavior of XTE~J1859$+$226. Variability appeared for this 
source whenever the power-law flux is higher than a threshold value. No other 
spectral variable shows a significant change between observations with and 
without variability \citep{Kalemci_tez}. The only caveat of this scenario is 
the resemblance of the variability properties of  XTE~J1859$+$226 to the IS 
observations of \gx\ and the 2000 outburst of \xf, rather than to the LS 
properties.

An increase in the power-law flux does not always correspond to the appearance
of variability. The 2000 outburst of \xf\ behaved differently; the rms 
amplitude of variability decreased as the the power-law flux increased
sharply. But in a couple of days after the first transition, another 
transition happened and strong variability that is usually associated with the
 LS appeared. The 1999 and the 2001 outbursts of \fu\ did not show a sharp 
change in any of the spectral properties at the transition 
(see Figs.~\ref{fig:evolmix1}~and~\ref{fig:evolmix2}), but both showed 
a change in the slope of the evolution of the power-law flux an observation 
earlier (see Fig.~\ref{fig:evolplf}). This suggests the possibility that for some 
cases the appearance of variability (in the case of \xf, variability associated 
with the LS) is delayed. The transition for these cases  may be happening slower 
than that of other cases, and during the restructuring of the accretion 
geometry, the coherence could be lost for all timescales and the timing 
signatures could be suppressed for a few days. 

\subsection{Evolution of spectral and temporal parameters}

Two important observations about the transitions to the LS can be made by 
analyzing the evolution of spectral parameters before the transition. First,
 the PLR increases for the majority of the outbursts. The corona (again 
assumed to be the source of the power-law flux) must dominate for the transition to
 happen, the PLR is greater than 0.65 for all sources after the transition to 
the LS. The PLR was as low as 0.45 for \maso\ when the variability was 
observed, but it never made the transition to the canonical LS, and returned 
back to the HS when the PLR dropped \citep{Kalemci_tez}. Second, the 
spectral indices for some of the sources show a step function-like decrease 
3-5 days  before the transition (see Fig.~\ref{fig:evolmix1}b). This may be 
interpreted as the time of spectral transition, which would indicate that the 
transition in timing properties lags the transition in spectral properties. Then, 
the change in  \ind\ may be a sign of changing the form of the corona as 
discussed in the previous section, and the lag between the change in the \ind\ 
and the appearance of variability is the time for the corona to reach a 
threshold volume. However, some sources do not show any drop in \ind, 
but still show a sharp change in temporal variability  (Fig.~\ref{fig:evolmix1}a), 
making the transitions much easier to identify.

After the transition, the evolution of the inner disk temperature, the diskbb
 flux, the power-law index and especially the characteristic frequencies of 
the variability are consistent with the idea of the inner disk retreat. In the 
basic formulation of \cite{Makishima86}, the optically thick disk has a 
temperature profile that decreases with increasing radius. Therefore, if the 
inner disk evaporates as part of the transition, it is expected that both 
\tin\ and diskbb flux decrease. Most sources show a decrease in diskbb flux, 
and at least four sources show a decrease in \tin. The drop of spectral index 
may also be a sign of increasing inner disk radius. As the disk is closer to 
the black hole, its temperature and flux are higher, causing effective cooling of 
the corona, increasing the spectral index \citep{Zdziarski02_2}. Four sources show
a decreasing \ind. It is intuitively expected that the characteristic 
frequencies decrease as the inner disk radius increases. The dynamical 
timescale (the fastest timescale) in the accretion disk is shorter close to 
the black hole, and it is expected that higher frequency variability is 
created in this region. As the inner disk radius retreats (or the inner disk 
evaporates) the dynamical timescale (at the inner edge) increases, and 
therefore the characteristic frequencies in the PSD decrease. The exact 
relation between the characteristic frequencies and the dynamical timescales 
is not clear yet. All sources show a decreasing behavior in terms of 
characteristic frequencies, except for one observation. The overall evolution 
of both spectral and temporal properties therefore indicates that the inner 
edge of the accretion disk retreats as the system progresses in the LS, in 
accordance with the model of \cite{Esin97}.  In the exceptional case, the QPO 
frequencies in the 2000 outburst of \xf\ first increased and then decreased 
after the transition (see Fig.~\ref{fig:evolcf}). This behavior is consistent 
with the prediction of the ``accretion ejection instability'' QPO model 
\citep{Tagger99} if the inner accretion disk is close to the marginally stable 
orbit during this observation. This possibility is discussed in detail in 
\cite{Rodriguez02b_inprep} and \cite{Kalemci_tez}. 

The interpretation of the behavior of the rms amplitudes is more complex. We 
showed the relation between the PLR and the rms amplitudes earlier (see 
Fig.~\ref{fig:simmulplot}). The ``Poissonic'' nature of the disk component 
causes a decrease in the rms amplitudes at lower energies due to the diskbb 
energy spectrum peaking at those energies. This might explain the increase and
 leveling of the rms amplitudes in \xs\ and the 2000 outburst of \xf. Although
 the rms amplitudes seem to be decreasing after the transition for the 1998 
outburst of \xf, it might be a result of not constraining the high frequency 
part of the PSD. The first observation requires two Lorentzians to fit, and 
results in a high PSD amplitude, whereas for the remaining observations, the 
second Lorentzian, although statistically not required, cannot be excluded 
\citep{Kalemci_tez}. The fits are better for \reso\ and it also 
shows a decreasing behavior. The interpretation of the rms amplitudes of
 \reso\ is complicated due to the high absorption column density. The Galactic
 ridge emission may also be affecting the rms amplitudes at low flux levels by
 supplying additional Poissonic emission. The rms amplitude of \gx\ increases
 after the transition although there is no diskbb flux in the 3-25 keV band. 
Overall, there is no definite trend of evolution for the rms amplitudes after
the transition.

\section{Summary}

An important goal of black hole binary research is to understand the 
accretion structure and nature of variability of these systems. This work 
addresses these goals by analyzing the X-ray temporal and spectral properties 
of a relatively large subset of GBHs during outburst decay. We characterize
the evolution of spectral and temporal properties before and after the 
transitions, and also work on the changes right at the transition to
understand the appearance of broad-band variability.

For this study, a total of seven \rxte\ sources in ten outbursts are analyzed. 
Thanks to our group's monitoring program, the coverage close to the transition 
is significantly improved, and for the first time allows us to determine the 
physical changes that drive the transitions in detail. The first problem we 
work on is the evolution of spectral parameters before the state transition and
 appearance of broad-band variability at the state transition. We show that 
the changes in variability properties (a sudden increase or decrease in the 
rms amplitude of variability) are sharper than changes in spectral properties, 
and it is easier to identify a transition with the temporal properties. A 
change in the spectral index \ind\ is shown to be the pre-cursor of the 
transition showing a decrease 3-5 days before for most of our sources. We also 
show that the PLR increases close to the transition for all sources. The hard
 power-law component must dominate the spectrum for the transition to happen. 
Very frequent monitoring observations (\wsim once a day) allow us to 
determine the power-law flux (and consequently the PLR) as the parameter 
showing a sharp change during the transitions which may indicate
a threshold volume for the corona for the appearance of variability.

We also investigate the evolution after the state transition. For most of the
cases, the \tin\ and the diskbb flux decrease after the transition, and they 
are unobservable with PCA within fifteen days of the transition for all 
outbursts. The characteristic frequencies of all except one observation of the 
2000 outburst of \xf\ decrease after the state transition. These spectral and 
temporal changes are consistent with the idea that the inner accretion disk 
retreats after the state transition. The rms amplitude of variability does not 
show any global trend with time after the transition, and its behavior is 
consistent with the emission from the soft component having no or very little 
temporal signature.


\acknowledgments 
E.K. acknowledges useful discussions with David Smith, Ali Alpar and \"Unal 
Ertan. The majority of this work was done at the Center for Astrophysics and
Space Sciences (CASS) at UCSD as part of E.K.'s dissertation. E.K. acknowledges 
the support of the Astrophysics Forum at Sabanc\i\ University where a portion 
of this work was prepared, and also acknowledges partial support of  
T\"UB\.ITAK. The authors would like to thank all scientists contributed to 
the T\"{u}bingen Timing Tools. J.A.T. acknowledges partial support from NASA grant 
NAG5-13055. K.P. was supported by grant Sta 173/25-1 and Sta 173/25-3 of the 
Deutsche Forschungsgemeinschaft. P.K. acknowledges partial support from NASA grant 
NAG5-7405. This work was also supported by NASA contract NAS5-30720.
  

\clearpage



\begin{sidewaystable}[t]
\scriptsize
\caption[Parameters for \xs]
{\label{table:1650} Paramters for \xs\footnote{\scriptsize This table is available only on-line 
as a machine-readable table.}
}
\begin{minipage}{\textwidth}
\begin{tabular}{l|c|ccccc|ccc} \hline \hline
Obs. & Date \footnote{MJD, Modified Julian Date, same for all Tables} & \ind & \tin & Power law Fl. & Dbb Fl. & PLR & rms (\%) & $\nu_{1}$\footnote{Lowest peak frequency, same for all Tables.} & QPO Freq.\footnote{Resonance frequency of the QPO if present, same for all Tables.} \\ \hline
1 &      52227.5 &  $4.390 \pm 0.343$ & $0.484 \pm 0.002$ & $2.400 \pm 0.264$ & $5.255 \pm 0.052$ & $0.313 \pm 0.030$ & $<$4\footnote{All upper limits are 2$\sigma$ in all
Tables.} & - & - \\
 2 &      52228.1 &  $4.026 \pm 0.271$ & $0.464 \pm 0.007$ & $2.211 \pm 0.221$ & $5.822 \pm 0.058$ & $0.275 \pm 0.024$ & $<4$ & - & - \\
 3 &      52230.4 &  $2.548 \pm 0.140$ & $0.487 \pm 0.008$ & $3.747 \pm 0.337$ & $5.080 \pm 0.050$ & $0.424 \pm 0.036$ & $<$4 & - & - \\
 4\footnote{The state transition happened between this observation and the next observation. The values in Figs.~\ref{fig:evolmix1},~\ref{fig:evolmix2},~\ref{fig:evolplf}
are normalized using parameters in this observation.Fig.~\ref{fig:evolcf} 
observations are normalized with respect to the next observation. Same for all 
tables.} &      52231.0 &  $2.415 \pm 0.038$ & $0.459 \pm 0.004$ & $4.524 \pm 0.135$ & $4.731 \pm 0.047$ & $0.488 \pm 0.015$ & $<$4 & - & - \\ \hline
5 &      52232.0 &  $2.379 \pm 0.019$ & $0.489 \pm 0.009$ & $12.76 \pm 0.127$ & $4.171 \pm 0.041$ & $0.753 \pm 0.009$ & $12.42 \pm 0.546$ & $4.162 \pm 0.498$ & - \\
 6 &      52233.2 &  $2.326 \pm 0.023$ & $0.443 \pm 0.016$ & $20.49 \pm 0.204$ & $2.221 \pm 0.022$ & $0.902 \pm 0.010$ & $19.01 \pm 1.672$ & $2.634 \pm 0.202$ & $8.734 \pm 0.219 $ \\
 7 &      52234.5 &  $2.093 \pm 0.011$ & $0.410 \pm 0.035$ & $30.62 \pm 0.306$ & $0.767 \pm 0.030$ & $0.975 \pm 0.012$ & $22.28 \pm 1.992$ & $1.274 \pm 0.123$ & $5.131 \pm 0.090 $ \\
8 &      52235.1 &  $2.096 \pm 0.009$ & $0.355 \pm 0.021$ & $29.70 \pm 0.297$ & $0.569 \pm 0.022$ & $0.981 \pm 0.012$ & $23.04 \pm 1.287$ & $1.408 \pm 0.089$ & $5.440 \pm 0.087 $ \\
 9 &      52236.1 &  $2.001 \pm 0.009$ & $0.374 \pm 0.033$ & $28.70 \pm 0.287$ & $0.517 \pm 0.036$ & $0.982 \pm 0.012$ & $25.16 \pm 1.614$ & $0.881 \pm 0.054$ & $4.655 \pm 0.072 $ \\
10 &      52236.9 &  $1.976 \pm 0.007$ & $0.321 \pm 0.011$ & $30.58 \pm 0.305$ & $0.271 \pm 0.027$ & $0.991 \pm 0.012$ & $24.75 \pm 0.330$ & $0.999 \pm 0.028$ & $3.799 \pm 0.034 $ \\
11 &      52237.9 &  $1.961 \pm 0.008$ & $0.306 \pm 0.006$ & $29.59 \pm 0.295$ & $0.144 \pm 0.014$ & $0.995 \pm 0.012$ & $25.45 \pm 2.449$ & $0.974 \pm 0.073$ & $3.830 \pm 0.077 $ \\ \hline
\end{tabular}
\end{minipage}
\end{sidewaystable}

\begin{sidewaystable}[t]
\scriptsize
\caption[Parameters for \xf\ in 1998]
{\label{table:1550x1} Paramters for \xf\ in 1998}
\begin{minipage}{\textwidth}
\begin{tabular}{l|c|ccccc|ccc} \hline \hline
Obs. & Date  & \ind & \tin & Power law Fl. & Dbb Fl. & PLR & rms (\%) & $\nu_{1}$ & QPO Freq. \\ \hline
1 &      1298.09 &  $2.439 \pm 0.134$ & $0.415 \pm 0.011$ & $0.867 \pm 0.065$ & $0.547 \pm 0.010$ & $0.613 \pm 0.048$ & $<$10 & - & - \\
 2 &      1299.34 &  $2.688 \pm 0.200$ & $0.392 \pm 0.011$ & $0.567 \pm 0.051$ & $0.389 \pm 0.007$ & $0.593 \pm 0.055$ &  $<$10 & - & - \\
 3 &      1302.46 &  $2.754 \pm 0.278$ & $0.394 \pm 0.023$ & $0.444 \pm 0.066$ & $0.219 \pm 0.004$ & $0.669 \pm 0.106$ &  $<$10 & - & - \\
 4 &      1303.39 &  $2.635 \pm 0.230$ & $0.379 \pm 0.027$ & $0.532 \pm 0.074$ & $0.187 \pm 0.005$ & $0.739 \pm 0.114$ &  $<$10 & - & - \\
 5 &      1305.12 &  $2.047 \pm 0.141$ & $0.464 \pm 0.030$ & $1.038 \pm 0.072$ & $0.228 \pm 0.006$ & $0.819 \pm 0.066$ & $<$10 & - & - \\ \hline
 6 &      1307.32 &  $1.968 \pm 0.031$ & $0.479 \pm 0.104$ & $3.746 \pm 0.074$ & $0.065 \pm 0.005$ & $0.982 \pm 0.024$ & $22.13 \pm 1.590$ & $1.175 \pm 0.210$ & - \\
 7 &      1309.72 &  $1.739 \pm 0.036$ & $0.595 \pm 0.084$ & $2.377 \pm 0.047$ & $0.092 \pm 0.007$ & $0.962 \pm 0.024$ & $17.74 \pm 0.600$ & $0.551 \pm 0.036$ & - \\
8 &      1311.52 &  $1.785 \pm 0.071$ & $0.518 \pm 0.073$ & $1.216 \pm 0.048$ & $0.069 \pm 0.006$ & $0.946 \pm 0.047$ & $14.81 \pm 1.590$ & $0.144 \pm 0.024$ & - \\
 9 &      1313.32 &  $1.773 \pm 0.088$ & $0.489 \pm 0.058$ & $0.861 \pm 0.051$ & $0.061 \pm 0.006$ & $0.933 \pm 0.068$ & $<$8 & - & - \\
10 &      1318.77 &  $1.815 \pm 0.132$ & $0.455 \pm 0.062$ & $0.491 \pm 0.039$ & $0.045 \pm 0.004$ & $0.916 \pm 0.089$ & $<$8 & - & - \\ \hline
\end{tabular}
\end{minipage}
\end{sidewaystable}

\begin{sidewaystable}[t]
\scriptsize
\caption[Parameters for \xf\ in 2000]
{\label{table:1550x2} Paramters for \xf\ in 2000}
\begin{minipage}{\textwidth}
\begin{tabular}{l|c|ccccc|ccc} \hline \hline
Obs. & Date  & \ind & \tin & Power law Fl. & Dbb Fl. & PLR & rms (\%) & $\nu_{1}$ & QPO Freq. \\ \hline
1 &      51667.7 &  $2.368 \pm 0.016$ & $0.798 \pm 0.010$ & $6.221 \pm 0.124$ & $4.539 \pm 0.045$ & $0.578 \pm 0.012$ & $12.50 \pm 0.470$ & $10.01 \pm 1.264$ & $9.030 \pm 0.240 $ \\
 2 &      51669.6 &  $2.386 \pm 0.014$ & $0.763 \pm 0.007$ & $5.170 \pm 0.103$ & $4.142 \pm 0.041$ & $0.555 \pm 0.012$ & $12.37 \pm 0.330$ & $9.571 \pm 0.894$ & $8.160 \pm 0.136 $ \\
 3 &      51670.6 &  $2.363 \pm 0.014$ & $0.756 \pm 0.010$ & $6.263 \pm 0.125$ & $3.307 \pm 0.033$ & $0.654 \pm 0.014$ & $13.12 \pm 0.280$ & $9.951 \pm 0.601$ & $9.320 \pm 0.088 $ \\
 4 &      51670.8 &  $2.362 \pm 0.015$ & $0.761 \pm 0.008$ & $5.467 \pm 0.109$ & $3.669 \pm 0.036$ & $0.598 \pm 0.013$ & $13.36 \pm 0.840$ & $9.615 \pm 0.932$ & $8.610 \pm 0.116 $ \\
 5 &      51671.4 &  $2.366 \pm 0.017$ & $0.748 \pm 0.008$ & $5.022 \pm 0.100$ & $3.550 \pm 0.035$ & $0.585 \pm 0.012$ & $13.06 \pm 0.690$ & $9.699 \pm 0.653$ & - \\
 6 &      51672.4 &  $2.342 \pm 0.015$ & $0.741 \pm 0.008$ & $5.031 \pm 0.100$ & $3.030 \pm 0.030$ & $0.624 \pm 0.013$ & $15.21 \pm 0.790$ & $8.906 \pm 0.768$ & $8.810 \pm 0.141 $ \\ \hline
 7 &      51673.0 &  $2.414 \pm 0.015$ & $0.761 \pm 0.014$ & $8.486 \pm 0.169$ & $2.857 \pm 0.028$ & $0.748 \pm 0.017$ & $7.230 \pm 0.260$ & $5.959 \pm 1.037$ & - \\
8 &      51673.4 &  $2.422 \pm 0.014$ & $0.759 \pm 0.012$ & $7.261 \pm 0.145$ & $2.798 \pm 0.027$ & $0.721 \pm 0.016$ & $8.440 \pm 0.290$ & $6.735 \pm 1.076$ & - \\
 9 &      51674.7 &  $2.331 \pm 0.014$ & $0.756 \pm 0.014$ & $8.471 \pm 0.169$ & $2.400 \pm 0.024$ & $0.779 \pm 0.017$ & $14.77 \pm 0.290$ & $1.019 \pm 0.257$ & $3.580 \pm 0.015 $ \\
10 &      51675.5 &  $2.284 \pm 0.012$ & $0.696 \pm 0.014$ & $7.164 \pm 0.143$ & $1.245 \pm 0.012$ & $0.851 \pm 0.019$ & $16.38 \pm 0.520$ & $1.794 \pm 0.137$ & $7.690 \pm 0.089 $ \\
11 &      51676.4 &  $2.275 \pm 0.014$ & $0.654 \pm 0.017$ & $6.346 \pm 0.126$ & $0.942 \pm 0.009$ & $0.870 \pm 0.020$ & $18.77 \pm 0.960$ & $2.425 \pm 0.429$ & $6.930 \pm 0.020 $ \\
12 &      51678.5 &  $2.116 \pm 0.011$ & $0.613 \pm 0.033$ & $6.233 \pm 0.124$ & $0.417 \pm 0.012$ & $0.937 \pm 0.022$ & $23.88 \pm 0.890$ & $1.148 \pm 0.092$ & $4.490 \pm 0.014 $ \\
13 &      51680.4 &  $2.143 \pm 0.009$ & $0.505 \pm 0.022$ & $5.489 \pm 0.164$ & $0.190 \pm 0.007$ & $0.966 \pm 0.035$ & $24.19 \pm 0.950$ & $1.014 \pm 0.083$ & $4.090 \pm 0.015 $ \\
14 &      51682.3 &  $1.883 \pm 0.007$ & $0.387 \pm 0.011$ & $4.520 \pm 0.135$ & $0.150 \pm 0.007$ & $0.967 \pm 0.035$ & $25.70 \pm 1.530$ & $0.558 \pm 0.029$ & $2.340 \pm 0.010 $ \\ \hline
\end{tabular}
\end{minipage}
\end{sidewaystable}

\begin{sidewaystable}[t]
\scriptsize
\caption[Parameters for \fu\ in 1998]
{\label{table:1630x1} Paramters for \fu\ in 1998}
\begin{minipage}{\textwidth}
\begin{tabular}{l|c|ccccc|ccc} \hline \hline
Obs. & Date  & \ind & \tin & Power law Fl. & Dbb Fl. & PLR & rms (\%) & $\nu_{1}$ & QPO Freq. \\ \hline
1 &     50937.630 &  $2.367 \pm 0.067$ & $0.764 \pm 0.014$ & $19.03 \pm 0.380$ & $10.05 \pm 0.100$ & $0.654 \pm 0.014$ & $3.000 \pm 0.200$ & - & - \\
 2 &    50939.070 &  $2.356 \pm 0.052$ & $0.738 \pm 0.012$ & $18.71 \pm 0.374$ & $8.851 \pm 0.088$ & $0.678 \pm 0.015$ & $3.500 \pm 0.200$ & - & - \\
 3 &     50942.010 &  $2.422 \pm 0.068$ & $0.758 \pm 0.011$ & $15.49 \pm 0.309$ & $10.24 \pm 0.102$ & $0.602 \pm 0.013$ & $2.500 \pm 0.300$ & - & - \\
 4 &    50945.860 &  $2.317 \pm 0.092$ & $0.711 \pm 0.016$ & $12.70 \pm 0.254$ & $6.887 \pm 0.068$ & $0.648 \pm 0.014$ & $4.000 \pm 0.500$ & - & - \\
 5 &    50949.740 &  $2.072 \pm 0.058$ & $0.601 \pm 0.012$ & $10.71 \pm 0.214$ & $3.826 \pm 0.038$ & $0.736 \pm 0.016$ & $2.000 \pm 0.800$ & - & - \\ \hline
 6 &    50951.670 &  $1.916 \pm 0.027$ & $0.461 \pm 0.017$ & $13.33 \pm 0.266$ & $1.201 \pm 0.024$ & $0.917 \pm 0.022$ & $10.20 \pm 0.600$ & $4.235 \pm 0.337$ & $3.390 \pm 0.008 $ \\
 7 &    50952.490 &  $1.858 \pm 0.032$ & $0.466 \pm 0.024$ & $9.092 \pm 0.181$ & $0.980 \pm 0.029$ & $0.902 \pm 0.022$ & $11.30 \pm 0.800$ & $3.540 \pm 0.247$ & $2.613 \pm 0.012 $ \\
8 &    50953.490 &  $1.798 \pm 0.027$ & $0.487 \pm 0.024$ & $10.64 \pm 0.212$ & $0.822 \pm 0.024$ & $0.928 \pm 0.022$ & $13.60 \pm 0.700$ & $3.140 \pm 0.128$ & $1.351 \pm 0.012 $ \\
 9 &    50956.960 &  $1.793 \pm 0.037$ & $0.459 \pm 0.026$ & $7.013 \pm 0.140$ & $0.510 \pm 0.025$ & $0.932 \pm 0.023$ & $15.90 \pm 1.100$ & $1.187 \pm 0.350$ & $0.430 \pm 0.006 $ \\
10 &    50958.960 &  $1.675 \pm 0.031$ & $0.486 \pm 0.025$ & $7.442 \pm 0.148$ & $0.565 \pm 0.028$ & $0.929 \pm 0.023$ & $16.10 \pm 1.500$ & $0.922 \pm 0.197$ & $0.365 \pm 0.011 $ \\
11 &    50962.960 &  $1.584 \pm 0.024$ & $0.452 \pm 0.029$ & $8.827 \pm 0.176$ & $0.406 \pm 0.040$ & $0.956 \pm 0.024$ & $17.30 \pm 0.800$ & $0.887 \pm 0.058$ & $0.228 \pm 0.003 $ \\ \hline
\end{tabular}
\end{minipage}
\end{sidewaystable}

\begin{sidewaystable}[t]
\scriptsize
\caption[Parameters for \fu\ in 1999]
{\label{table:1630x2} Paramters for \fu\ in 1999}
\begin{minipage}{\textwidth}
\begin{tabular}{l|c|ccccc|ccc} \hline \hline
Obs. & Date  & \ind & \tin & Power law Fl. & Dbb Fl. & PLR & rms (\%) & $\nu_{1}$ & QPO Freq. \\ \hline
1 &      51388.3 &  $2.060 \pm 0.036$ & $0.511 \pm 0.014$ & $9.891 \pm 0.197$ & $2.414 \pm 0.048$ & $0.803 \pm 0.019$ & $<$2 & - & - \\
 2 &      51389.1 &  $1.987 \pm 0.028$ & $0.519 \pm 0.011$ & $10.09 \pm 0.201$ & $2.274 \pm 0.045$ & $0.816 \pm 0.019$ & $<$2 & - & - \\
 3 &      51390.1 &  $2.036 \pm 0.032$ & $0.529 \pm 0.014$ & $8.340 \pm 0.166$ & $2.140 \pm 0.042$ & $0.795 \pm 0.019$ & $<$2 & - & - \\
 4 &      51391.4 &  $2.014 \pm 0.032$ & $0.500 \pm 0.016$ & $7.680 \pm 0.153$ & $1.557 \pm 0.046$ & $0.831 \pm 0.020$ & $<$2 & - & - \\
 5 &      51393.2 &  $1.834 \pm 0.024$ & $0.464 \pm 0.014$ & $8.301 \pm 0.166$ & $1.220 \pm 0.061$ & $0.871 \pm 0.022$ & $<$2 & - & - \\ \hline
 6 &      51394.8 &  $1.845 \pm 0.024$ & $0.469 \pm 0.027$ & $7.294 \pm 0.145$ & $0.796 \pm 0.055$ & $0.901 \pm 0.023$ & $12.50 \pm 2.850$ & $4.090 \pm 0.089$ & $1.820 \pm 0.004 $ \\
 7 &      51395.5 &  $1.881 \pm 0.031$ & $0.429 \pm 0.032$ & $6.033 \pm 0.120$ & $0.694 \pm 0.062$ & $0.896 \pm 0.024$ & $8.553 \pm 1.000$ & $0.603 \pm 0.517$ & $0.782 \pm 0.083 $ \\
8 &      51397.2 &  $2.041 \pm 0.076$ & $0.424 \pm 0.058$ & $3.483 \pm 0.104$ & $0.438 \pm 0.043$ & $0.888 \pm 0.035$ & $<$4 & - & - \\
 9 &      51398.2 &  $1.802 \pm 0.093$ & $0.378 \pm 0.041$ & $3.854 \pm 0.154$ & $0.351 \pm 0.035$ & $0.916 \pm 0.046$ & $<$4 & - & - \\
10 &      51400.1 &  $2.048 \pm 0.061$ & $0.425 \pm 0.053$ & $2.852 \pm 0.142$ & $0.251 \pm 0.025$ & $0.919 \pm 0.057$ & $<$4 & - & - \\ \hline
\end{tabular}
\end{minipage}
\end{sidewaystable}

\begin{sidewaystable}[t]
\scriptsize
\caption[Parameters for \fu\ in 2001]
{\label{table:1630x3} Paramters for \fu\ in 2001}
\begin{minipage}{\textwidth}
\begin{tabular}{l|c|ccccc|ccc} \hline \hline
Obs. & Date  & \ind & \tin & Power law Fl. & Dbb Fl. & PLR & rms (\%) & $\nu_{1}$ & QPO Freq. \\ \hline
1 &      52048.0 &  $2.219 \pm 0.039$ & $0.784 \pm 0.012$ & $18.01 \pm 0.360$ & $15.18 \pm 0.151$ & $0.542 \pm 0.011$ & $<$2 & - & - \\
 2 &      52049.8 &  $2.236 \pm 0.034$ & $0.702 \pm 0.012$ & $15.27 \pm 0.305$ & $10.19 \pm 0.101$ & $0.599 \pm 0.013$ & $<$2 & - & - \\
 3 &      52051.6 &  $2.260 \pm 0.029$ & $0.674 \pm 0.012$ & $14.32 \pm 0.286$ & $7.206 \pm 0.072$ & $0.665 \pm 0.014$ & $<$2 & - & - \\
 4 &      52053.8 &  $2.095 \pm 0.020$ & $0.592 \pm 0.011$ & $13.25 \pm 0.265$ & $4.642 \pm 0.046$ & $0.740 \pm 0.016$ & $<$2 & - & - \\
 5 &      52055.6 &  $2.230 \pm 0.038$ & $0.542 \pm 0.014$ & $9.742 \pm 0.194$ & $3.458 \pm 0.069$ & $0.738 \pm 0.017$ & $<$2 & - & - \\
 6 &      52058.0 &  $1.970 \pm 0.030$ & $0.460 \pm 0.017$ & $10.29 \pm 0.205$ & $1.760 \pm 0.052$ & $0.853 \pm 0.021$ & $<$2 & - & - \\ \hline
 7 &      52059.4 &  $1.575 \pm 0.010$ & $0.401 \pm 0.016$ & $11.06 \pm 0.221$ & $0.895 \pm 0.062$ & $0.925 \pm 0.024$ & $11.72 \pm 2.730$ & $2.520 \pm 1.070$ & $1.396 \pm 0.042 $ \\
8 &      52061.6 &  $1.683 \pm 0.030$ & $0.463 \pm 0.023$ & $4.766 \pm 0.142$ & $0.767 \pm 0.053$ & $0.861 \pm 0.033$ & $9.323 \pm 3.150$ & $0.297 \pm 0.069$ & - \\
 9 &      52063.6 &  $1.751 \pm 0.026$ & $0.444 \pm 0.014$ & $3.961 \pm 0.158$ & $0.590 \pm 0.047$ & $0.870 \pm 0.044$ & $<$5 & - & - \\
10 &      52065.9 &  $1.694 \pm 0.021$ & $0.430 \pm 0.025$ & $3.880 \pm 0.194$ & $0.491 \pm 0.049$ & $0.887 \pm 0.056$ & $<$5 & - & - \\ \hline
\end{tabular}
\end{minipage}
\end{sidewaystable}

\begin{sidewaystable}[t]
\scriptsize
\caption[Parameters for \reso]
{\label{table:1748} Paramters for \reso}
\begin{minipage}{\textwidth}
\begin{tabular}{l|c|ccccc|ccc} \hline \hline
Obs. & Date & \ind & \tin & Power law Fl. & Dbb Fl. & PLR & rms (\%) & $\nu_{1}$ & QPO Freq. \\ \hline
1 &      51002.7 &  $2.643 \pm 0.031$ & $0.892 \pm 0.015$ & $17.73 \pm 0.354$ & $9.967 \pm 0.099$ & $0.640 \pm 0.014$ & $<$2 & - & - \\
 2 &      51007.3 &  $1.967 \pm 0.017$ & $0.437 \pm 0.015$ & $19.62 \pm 0.392$ & $2.496 \pm 0.074$ & $0.887 \pm 0.021$ & $4.250 \pm 0.250$ & - & - \\ \hline
 3 &      51012.2 &  $1.874 \pm 0.005$ & $0.396 \pm 0.030$ & $10.07 \pm 0.201$ & $0.253 \pm 0.025$ & $0.975 \pm 0.024$ & $21.14 \pm 3.160$ & $1.104 \pm 0.186$ & - \\
 4 &      51024.4 &  $1.828 \pm 0.013$ & $0.399 \pm 0.019$ & $8.879 \pm 0.177$ & $0.270 \pm 0.027$ & $0.970 \pm 0.024$ & $20.71 \pm 0.370$ & $0.220 \pm 0.014$ & - \\ \hline
\end{tabular}
\end{minipage}
\end{sidewaystable}

\begin{sidewaystable}[t]
\scriptsize
\caption[Parameters for GRO~J1655$-$40]
{\label{table:1655} Paramters for GRO~J1655$-$40}
\begin{minipage}{\textwidth}
\begin{tabular}{l|c|ccccc|ccc} \hline \hline
Obs. & Date & \ind & \tin & Power law Fl. & Dbb Fl. & PLR & rms (\%) & $\nu_{1}$ & QPO Freq. \\ \hline
1 &      50658.4 &  $2.106 \pm 0.029$ & $0.898 \pm 0.004$ & $16.25 \pm 0.325$ & $57.22 \pm 0.572$ & $0.221 \pm 0.004$ & $1.800 \pm 0.100$ & - & - \\
 2 &      50663.7 &  $2.392 \pm 0.092$ & $0.795 \pm 0.003$ & $9.031 \pm 0.180$ & $29.68 \pm 0.296$ & $0.233 \pm 0.004$ & $<$2 & - & - \\ \hline
 3 &      50674.4 &  $1.948 \pm 0.010$ & $0.610 \pm 0.027$ & $27.59 \pm 0.551$ & $1.220 \pm 0.036$ & $0.957 \pm 0.023$ & $22.66 \pm 0.770$ & $1.514 \pm 0.077$ & $6.457 \pm 0.020 $ \\
 4 &      50678.6 &  $1.670 \pm 0.015$ & $0.629 \pm 0.077$ & $9.284 \pm 0.278$ & $0.203 \pm 0.020$ & $0.978 \pm 0.036$ & $24.34 \pm 2.390$ & $0.313 \pm 0.027$ & $0.785 \pm 0.011 $ \\
 5 &      50685.4 &  $1.751 \pm 0.057$ & $0.485 \pm 0.026$ & $0.971 \pm 0.058$ & $0.072 \pm 0.007$ & $0.930 \pm 0.068$ & $<$5 & - & - \\ \hline
\end{tabular}
\end{minipage}
\end{sidewaystable}

\end{document}